\documentstyle[aps,preprint,
]{revtex} 
\tighten
\author{F.N. Braun and C. Viney}
\address{Department of Chemistry, Heriot-Watt University,
Edinburgh EH14 4AS, United Kingdom}
\title{Self-assembly 
in the major ampullate gland of Nephila clavipes}
\begin{document}
\maketitle
\begin{abstract}
We present a tentative  interpretation of the
origin of nematic liquid crystalline order exhibited  by  
dragline silk fibroin solutions collected from the spider 
{\it Nephila clavipes}. Liquid crystallinity is thought to
confer certain rheological properties on the fibroin solution
which are exploited during the dragline spinning process.
We show that the feasibility of liquid crystallinity
under physiological conditions depends critically on 
parameters characterising the amino-acid sequence of the fibroin molecules. 

\end{abstract}
\bigskip

``Dragline" silk secreted by the major ampullate gland of the orb-weaver spider
{\it Nephila clavipes} is one of the best-characterised natural
silks, notable in respect of its singular mechanical properties \cite{Zemlin}.
The spinning process itself, however, i.e., the structural 
transition undergone by fibroins initially dispersed in the aqueous 
environment of the gland, culminating with the extrusion of
water-insoluble fiber, remains relatively poorly understood \cite{Guehrs}. 
Recent {\it in vitro} experiments \cite{Kerkam}
have suggested that the solution in the gland enters 
a nematic (or twisted nematic) liquid crystalline state, which might
play a rheological role in the spinning process
and hence affect the mechanical quality of the extruded fiber \cite{Vollrath}.

Interestingly, Gatesy {\it et al.}\cite{Gatesy} have argued
that, in the evolutionary context, mechanical performance of natural silks 
exerts a particularly acute selection pressure  
on the amino-acid sequence of the constituent fibroin molecules. A stabilizing
source of selection pressure appears necessary to explain
the remarkably long evolutionary timescale ($\sim$125 million years) 
over which the characteristic repeated-motif sequences
of orb-weaving spiders have remained essentially unchanged.
In this regard, correspondence between sequence and the 
ability to form a liquid crystalline phase in the gland,
affecting spinning conditions, and hence dragline quality,
would present a relevant perspective. 

Our objective here is to tentatively establish such a connection,
addressing at the molecular level how nematic order emerges 
in the gland, and is controlled by sequence-related parameters. 
We develop from the premise that the dragline fibroin solution 
belongs to a general class of fibrillizing globular protein systems of which 
hemoglobin S is a well-known example\cite{Perutz}. 
The globules of these systems assemble reversibly into supramolecular 
rod-like structures, which, beyond  a critical axial ratio, undergo a 
nematic ordering transition. 
\medskip

The details of the assembly process are as follows. 
We identify three generic sources of protein-protein interaction 
contributing to an effective binding potential between fibroins.
The principal contribution derives from the
free energy of transfer $\Delta F$ of 
nonpolar amino acid from a hydrophobic environment to water -
the so-called hydrophobic effect.
According to data collated and interpreted by Dill {\it et al.} 
\cite{Dill}, $\Delta F$  has a magnitude of approximately 2 kcal/mol,
increasing with temperature up to a maximum
at 60-80$^o$C:
\[
\Delta F(T)=\Delta H^0+\Delta  C_p(T-T_0)-
T[\Delta S^0+\Delta  C_p\ln(T/T_0)],
\]
where $\Delta H^0=0$ and $\Delta S^0=-6.7$ cal/K/mol are
the enthalpy and entropy of transfer at room
temperature $T_0=298K$, and $\Delta  C_p=55$ cal/K/mol is the  heat capacity.

If we assume that each fibroin monomer is in a dense-packed
roughly spherical globular state (radius $r$), the hydrophobic effect
generates an effective fibroin-fibroin binding  potential 
\begin{equation}
\gamma_h\simeq n_h\Delta F(T)\sim 10 {\,\rm cal/mol/\AA^2},
\end{equation}
per unit area of surface buried at the binding contact,
where, with $a\sim 1$nm a characteristic residue lengthscale, 
$n_h\sim a^{-2}$ is the area density of
hydrophobic residues exposed at the globule surface.

Dispersion forces likewise favour association, but contribute
a much smaller term  \cite{Hunter}
\begin{equation}
\gamma_d= \frac{A}{24\pi a^2}< {1\rm cal/mol/\AA^2},
\end{equation}
where the Hamaker constant $A$ gives the strength of 
water-mediated dispersion interaction between amino-acids residues,
This estimate assumes $A\simeq 500$ cal/mol, as for typical alkane chains in water \cite{Hunter}.

Thirdly, the silk fibroin sequence \cite{Xu} features a number 
$N_{\rm Arg}\sim 20$ of arginine residues, which carry a positive 
charge at normal pH \cite{Bryn}.  These residues generate an electrostatic 
double layer effect which we can examine, to a first approximation, 
by treating fibroin globules as conducting spheres each
carrying a charge $\alpha N_{\rm Arg}$, where
the `degree of ionization' $\alpha$ reflects
the extent to which the solvent is able to penetrate into the globule interior. 

The free energy of formation of the double layer is\cite{Stigter}
\[
-\frac{\alpha^2 N_{\rm Arg}^2e^2}{8\pi\epsilon r}\left(\frac{\xi}{r}\right)
\hspace{.3in} (r>>\xi),
\]
where $\xi$ is the Debye screening length, 
and $\epsilon\simeq 7\times 10^{-11}$ F/m is the dielectric constant of water.
The limit $r>>\xi$ is appropriate at salt concentrations
beyond the order of $10^{-2}M$, for which the Debye length is 
comparable with the amino-acid residue dimension, $\xi\sim a$.

A negative  contribution to the binding potential per unit area 
of fibroin-fibroin interface follows 
from the unfavourable cost of breaking up this double layer,
\begin{equation}
\gamma_e \simeq -
\frac{\alpha^2 N_{\rm Arg}^2e^2}{32\pi^2\epsilon r^3}\left(\frac{\xi}{r}\right)
\label{e1}
\end{equation}
We observe that binding is only feasible if the degree of ionization
is very low, since otherwise only three or four arginine residues 
would be sufficient to generate magnitudes 
$\gamma_e \sim -10 {\,\rm cal/mol/\AA^2}$ comparable 
with the hydrophobic term, hence preventing association. 

The buried surface area of a fibroin embedded within a supramolecular mesogen
scales with the fibroin size as $r^2$, yielding
for the net binding free energy due to the respective contributions 
\begin{equation}
E_b= {\rm const} \times r^2(\gamma_h+\gamma_d+\gamma_e),
\label{eq}
\end{equation}
where the constant is a geometrical
factor of order $\sim 4\pi$ which depends on the morphology of
the supramolecular structure.  In a `close-packed' estimate for $r$, we have
$r/a =(3N/4\pi)^{1/3}\simeq 6$,
where $N\sim700$ is the total number of residues per globule 
according to the sequence data \cite{Xu}.

The binding potential is attractive when 
the  hydrophobic term dominates over the electrostatic repulsion, 
yielding magnitudes $E_b\sim 10 RT$ cal/mol.
In order to relate $E_b$ to mesogen assembly,
we adopt a well-known approximate result of
generalized Flory-Huggins lattice statistics \cite{Cates},
giving for the mean axial ratio of rods reversibly
assembled from associating monomers at volume fraction $\Phi$ \cite{time},
\begin{equation}
x(\Phi,T) =s^{-3}\Phi^{1/2} \exp(E_b/2RT).
\label{ax}
\end{equation}

Here $s$ specifies the number of monomer `strands' in the rod
cross-section.
In the example of hemoglobin S fibrils, $s=14$ entwined strands have been
identified from crystallographic data \cite{Perutz}.
Unfortunately, in the absence of similar data
for silk fibrils in the major ampullate gland, $s$ is an unknown parameter
in the present context \cite{Aggeli}. 

Nematic-isotropic coexistence is naively calculated
in Fig.1 (for a hypothetical choice of $s$), 
by solving Onsager-like criteria \cite{Onsager}
defining respectively the isotropic (I) and nematic (I) nodal lines 
bounding the biphasic region,
\begin{equation}
\left.x(\Phi,T)\right|_I=5/\Phi \hspace{.4in}\left.x(\Phi,T)\right|_N=8/\Phi. 
\label{tr}
\end{equation}

In this type of calculation, we neglect thermodynamic coupling between nematic 
ordering and mesogen assembly \cite{Gelbart}, and related 
coupling to polydispersity \cite{poly} and flexibility \cite{flex}
of the mesogens. 
Anisotropic electrostatic \cite{charge} and dispersion force \cite{disp}
contributions to the effective mesogen-mesogen 
interaction, which is purely steric in the Onsager approach, 
are similarly neglected.
\bigskip

In summary, we have shown that it is feasible for silk solutions to exhibit 
nematic liquid crystalline order along broadly similar lines to
the hemoglobin S mechanism.
The hydrophobic effect drives  thermodynamically reversible assembly 
of supramolecular mesogens,
which are capable of ordering according to an Onsager-like criterion.  

This interpretation remains only tentative, insofar
as we do not have cystallographic data to hand against which to elucidate 
structural features of fibroin monomers and fibrils which are implicit 
to the approach.
In particular, values for $s$, characterising mesogen morphology,
and $n_h$, specifying the hydrophobic surface topology of fibroin monomers,
are unknown. 

The strong sensitivity of the model to these parameters, along with
unspecified geometrical constants,
renders the semi-quantitative phase diagram calculated in the figure 
somewhat academic.
However, the reentrant nature of the phase diagram, reflecting
the temperature dependence of the hydrophobic effect,
presents a strong qualitative signature which
might be looked for experimentally.

There is some scope for comment on the suggestion that 
spiders regulate their silk gland {\it in vivo} by pumping in 
additional protons and salts (see \cite{Chen} and references therein).
According to our Eqns (\ref{e1}-\ref{eq}), there is a 
strengthening of the net 
fibroin-fibroin binding potential in response to added salt, 
due to increased screening of the electrostatic double layer repulsion,
which should favour mesogen assembly and the onset of a nematic phase.
This is in line with very recent observations of 
Chen {\it et al.}\cite{Chen}, who report that nanofibril formation 
can be induced in {\it Nephila senegalensis} major ampullate 
fibroin solutions by the addition of KCl. These authors also 
observe a marked transition in the rheological character
of the solution occuring between pH 6.4 and 6.8, which provides us
with some justification for the significance attributed in the model
to histidine residues (pK value 6.5).

Finally, the model affords some insight into how sequence mutations might 
be expected to interfere with liquid crystallinity of the fibroin solution.
The strong shift of the biphasic region calculated in the figure, for example, 
roughly reflects a single site hydrophobic$\rightarrow$ polar mutation
in the surface of the interacting globules.
We infer that mutations are in general easily capable of 
disrupting the dragline spinning process
by rendering the nematic phase physiologically inaccessible.
Arguably, this has attendant consequences for the evolution of the fibroins.
Sequence mutations which interfere with 
liquid crystallinity are discouraged by selection pressure, 
since by disrupting the spinning process  they
would also degrade mechanical performance of the extruded dragline.

\begin{figure}
\bigskip
\caption{Nematic-isotropic biphasic region of the major ampullate gland 
phase diagram, from Eqns (\ref{ax}-\ref{tr}) with $E_b =n_h\Delta F(T)$.  
The dispersion force and electrostatic contributions to $E_b$ as presented 
in the text are relatively weak and have been neglected.
Substituted parameters are $s=8$, $r/a=5$,
and the thermodynamic parameters for $\Delta F(T)$ quoted in the  text.
We contrast the biphasic region for $n_h=.25a^{-2}$ (solid lines)
with $n_h=.23a^{-2}$ (dashed lines).
This parameter shift is chosen to roughly reflect a single site 
hydrophobic$\rightarrow$ polar mutation
in the surface of the interacting globules.
}
\end{figure}

\medskip


\begin{references}
\bibitem{Zemlin}J.C. Zemlin, {\it A study of the mechanical behaviour
of spider silks}, U.S. Army Natick Lab. Technical Report 69-29-CM (1968);
M.W. Denny, J. Exp. Biol. {\bf 65} 483 (1976).
\bibitem{Guehrs}K.H. G\"uhrs, K. Weisshart and F. Grosse,
Rev. Mol. Biotechnology {\bf 74} 121 (2000). 
\bibitem{Kerkam}K. Kerkam, C.Viney, D. Kaplan and S. Lombardi,
Nature {\bf 349} 596 (1991); P.J.Willcox and S.P. Gido, Macromolecules {\bf 29} 5106 (1996);
C. Viney, Supramolecular Science {\bf 4} 75 (1997);D.P. Knight and
F. Vollrath, Proc. R. Soc. Lond. B {\bf 266} 519 (1999).
\bibitem{Vollrath} F. Vollrath and D.P. Knight, Nature {\bf 410} 541 (2001).
\bibitem{Gatesy}J. Gatesy, C. Hayashi, D. Motriuk, J. Woods and R. Lewis,
Science {\bf 291} 2603 (2001); C. Hayashi and R.V. Lewis, Science {\bf
 287} 1477 (2000).
\bibitem{Perutz}H.F. Perutz, A. Liquori, F. Eirich, Nature {\bf 167} 929 (1951);A.C. Allison, Biochem. J. {\bf 65} 212 (1957);
W.A. Eaton and J. Hofrichter, Adv. Protein Chem. {\bf 40} 63 (1990).
\bibitem{Dill}K.A. Dill, D.O.V. Alonso and K. Hutchinson,
Biochemistry {\bf 28} 5439 (1989).
\bibitem{Hunter} this result is from the standard Hamaker theory,
see e.g. R.J. Hunter, {\it Foundations of Colloid Science,
vol.1} (Oxford Science, 1995).
\bibitem{Xu} M. Xu and R.V. Lewis, Proc. Natl.  Acad. Sci. USA, {\bf 87} 7120 (1990). 
\bibitem{Bryn}J.D. Bryngelson and E.M. Billings, in {\it Physics of
Biological Systems} (Springer, 1997, Berlin)
\bibitem{Stigter}The result quoted here 
involves the Debye-H\"uckel  approximation, 
which is subject to the conditon $Z^2e^2/(4\pi\epsilon r)<< k_BT$.
Details of the Debye-H\"uckel and more sophisticated 
approaches to solving the Poisson-Boltzmann equation 
around protein globules are discussed by
D. Stigter and K.A. Dill, Biochemistry {\bf 29} 1262 (1990), 
D. Stigter, D.O.V. Alonso and K.A. Dill, Proc. Natl. Acad. Sci. USA
{\bf 88} 4176 (1991).
\bibitem{Cates}M.E. Cates and S.J. Candau, J. Phys: Cond. Mat. {\bf 2} 6869 (1990)
\bibitem{time}
The timescale governing this assembly process is critical
to the functional context of the major ampullate gland. 
For diffusion-limited association kinetics we have
by a well-known textbook argument $\tau\sim (\eta r^3)/(\Phi k_BT)$
where $\eta\sim$10 Poise is the viscosity  of water, from 
which we estimate a functionally plausible $\tau\sim$1ms.
\bibitem{Aggeli}
The  high propensity of silk fibroins towards 
$\beta$-sheet formation (see \cite{Xu})
is relevant here. A recent study of self-assembling peptide
solutions suggests that $s$ is constrained by $\beta$-sheet induced
chirality of the monomer units:
A. Aggeli, I.A. Nyrkova, M. Bell, R. Harding,
L. Carrick, T.C.B. McLeish, A.N. Semenov and N. Boden,
Proc. Natl. Acad. Sci. USA {\bf 98} 11857 (2001).
\bibitem{Onsager}
The Onsager theory interprets athermal nematic ordering of 
monodisperse hard rods as the result of
competition between orientiational entropy and orientation-dependent
excluded volume. The general form of the coexistence
criteria quoted here is also arrived at in a lattice approach due to Flory, 
although the two theories predict different values of the constant factors 
which we have taken as 5 and 8. See 
L. Onsager, Ann. N.Y. Acad. Sci. {\bf 51} 627 (1950);
P.J. Flory, Proc. Roy. Soc. London Ser. A {\bf 234} 73 (1956).
\bibitem{Gelbart}W.M. Gelbart, W.E. McMullen and A. Ben-Shaul,
J. Phys. (Paris) {\bf 46} 1137 (1985).
\bibitem{poly}W.E. McMullen, W.M. Gelbart, A. Ben-Shaul,
J. Chem. Phys. {\bf 82} 5616 (1985); 
J. Herzfeld, J. Chem. Phys. {\bf 88} 2776 (1988). 
\bibitem{flex}T. Odijk, J. Phys. (Paris) {\bf 48} 125 (1987);
R. Hentschke and J. Herzfeld, Phys. Rev. A {\bf 44} 1148 (1991).
\bibitem{charge}A. Stroobants, H.N.W. Lekkerkerker, T. Odijk,
Macromolecules {\bf 19} 2233 (1986).
\bibitem{disp}W. Maier and A. Saupe, Z. Naturforsch. {\bf 13a} 564 (1958).
\bibitem{Chen}X. Chen, D.P. Knight and F. Vollrath, Biomacromol. {\bf 3}
644 (2002).
\end{references}
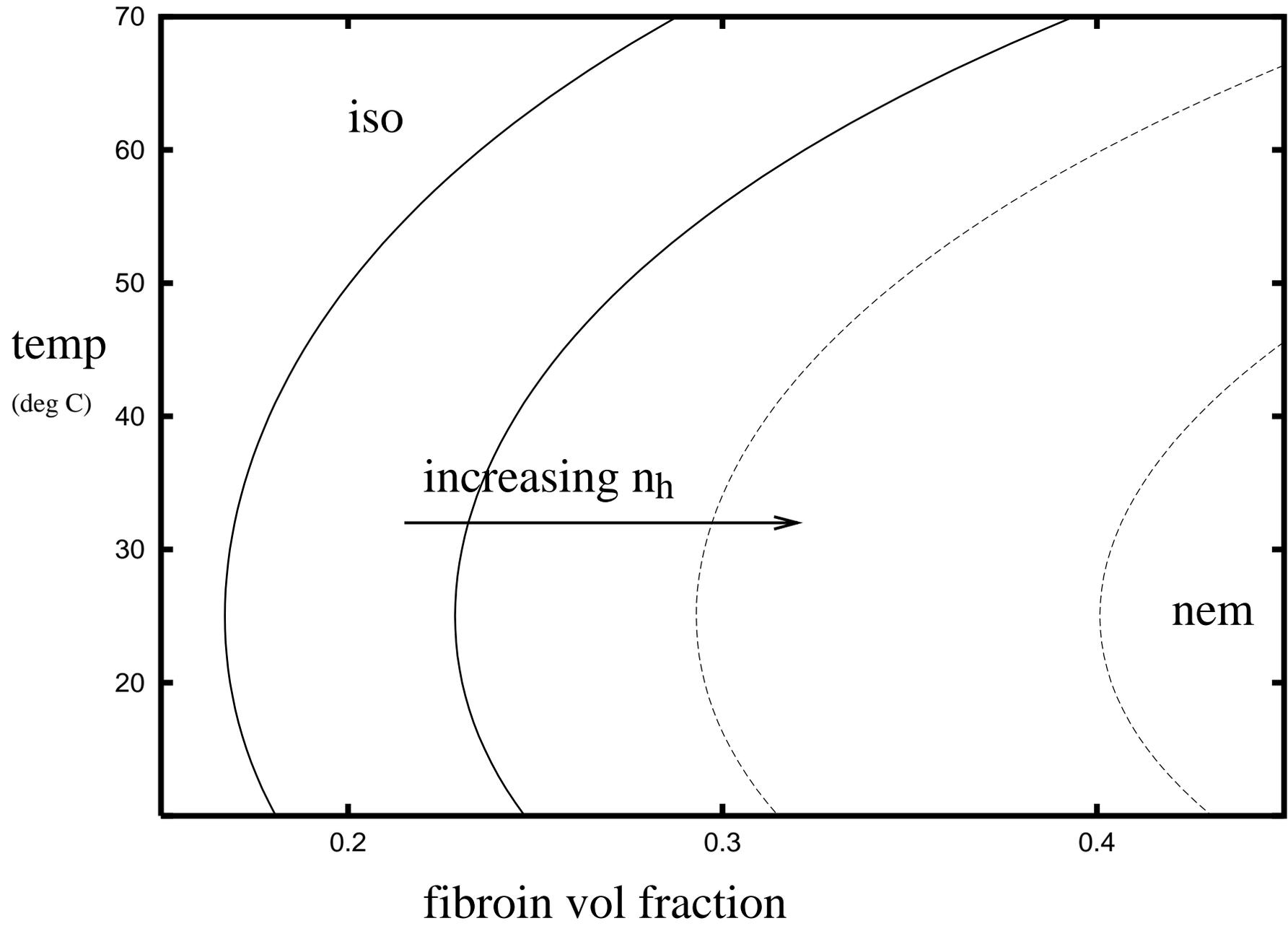
\end{document}